\documentclass[12pt,preprint]{aastex}

\shorttitle{$>$10~TeV $\gamma$-rays from Mrk~421}
\shortauthors{Okumura et al.}

\begin{document}

\title{Observation of gamma-rays greater than 10~TeV 
from Markarian~421}

\author{
K.~Okumura\altaffilmark{1},
A.~Asahara\altaffilmark{2},
G.V.~Bicknell\altaffilmark{3},
P.G.~Edwards\altaffilmark{4},
R.~Enomoto\altaffilmark{1},
S.~Gunji\altaffilmark{5},
S.~Hara\altaffilmark{2,6},
T.~Hara\altaffilmark{7},
S.~Hayashi\altaffilmark{8},
C.~Itoh\altaffilmark{9},
S.~Kabuki\altaffilmark{1},
F.~Kajino\altaffilmark{8},
H.~Katagiri\altaffilmark{1},
J.~Kataoka\altaffilmark{6},
A.~Kawachi\altaffilmark{1},
T.~Kifune\altaffilmark{10},
H.~Kubo\altaffilmark{2},
J.~Kushida\altaffilmark{2,6},
S.~Maeda\altaffilmark{8},
A.~Maeshiro\altaffilmark{8},
Y.~Matsubara\altaffilmark{11},
Y.~Mizumoto\altaffilmark{12},
M.~Mori\altaffilmark{1},
M.~Moriya\altaffilmark{6},
H.~Muraishi\altaffilmark{13},
Y.~Muraki\altaffilmark{11},
T.~Naito\altaffilmark{7},
T.~Nakase\altaffilmark{14},
K.~Nishijima\altaffilmark{14},
M.~Ohishi\altaffilmark{1},
J.R.~Patterson\altaffilmark{15},
K.~Sakurazawa\altaffilmark{6},
R.~Suzuki\altaffilmark{1},
D.L.~Swaby\altaffilmark{15},
K.~Takano\altaffilmark{6},
T.~Takano\altaffilmark{5},
T.~Tanimori\altaffilmark{2},
F.~Tokanai\altaffilmark{5},
K.~Tsuchiya\altaffilmark{1},
H.~Tsunoo\altaffilmark{1},
K.~Uruma\altaffilmark{14},
A.~Watanabe\altaffilmark{5},
S.~Yanagita\altaffilmark{9},
T.~Yoshida\altaffilmark{9},
and T.~Yoshikoshi\altaffilmark{16}
}

\altaffiltext{1}{Institute for Cosmic Ray Research, University of Tokyo,
Chiba 277-8582, Japan}
\altaffiltext{2}{Department of Physics, Kyoto University, Kyoto 606-8502, Japan}
\altaffiltext{3}{MSSSO, Australian National University, ACT 2611, Australia}
\altaffiltext{4}{Institute of Space and Astronautical Science, Kanagawa
229-8510, Japan}

\altaffiltext{5}{Department of Physics, Yamagata University, Yamagata 990-8560, Japan}
\altaffiltext{6}{Department of Physics, Tokyo Institute of Technology,
Tokyo 152-8551, Japan}
\altaffiltext{7}{Faculty of Management Information, Yamanashi Gakuin
University, Yamanashi 400-8575, Japan}
\altaffiltext{8}{Department of Physics, Konan University, Hyogo 658-8501, Japan}
\altaffiltext{9}{Faculty of Science, Ibaraki University, Ibaraki 310-8512, Japan}
\altaffiltext{10}{Faculty of Engineering, Shinshu University, Nagano 380-8553, Japan}
\altaffiltext{11}{STE Laboratory, Nagoya University, Aichi 464-8601, Japan}
\altaffiltext{12}{National Astronomical Observatory of Japan, Tokyo 181-8588, Japan}
\altaffiltext{13}{Ibaraki Prefectural University of Health Sciences, Ibaraki 300-0394, Japan}
\altaffiltext{14}{Department of Physics, Tokai University, Kanagawa 259-1292, Japan}
\altaffiltext{15}{Department of Physics and Math. Physics, University of
Adelaide, SA 5005, Australia}
\altaffiltext{16}{Department of Physics, Osaka City University, Osaka 558-858, Japan}

\begin{abstract}

We have observed Markarian~421
in January and March 2001 
with the CANGAROO-II imaging Cherenkov telescope
during an extraordinarily high state
at TeV energies.
From 14~hours observations at very large zenith angles,
$\sim$70$^\circ$, 
a signal of 298\,$\pm$\,52 gamma-ray--like events (5.7~$\sigma$) 
was detected at $E>10$~TeV, 
where a higher sensitivity is achieved than those of usual
observations near the zenith,
owing to a greatly increased collecting area.
Under the assumption of an intrinsic power-law spectrum,  
we derived a differential energy spectrum 
 $dN/dE = (3.3\,\pm\,0.9_{stat.}\,\pm\,0.3_{syst.})\times10^{-13}~(E/10~\mbox{TeV})^{-(4.0\,^{+0.9}_{-0.6}\,_{stat.}\,\pm\,0.3_{syst.})}$  ph./cm$^2$/sec/TeV,
which is steeper than those previously measured around 1~TeV,
and supports the evidence for a cutoff in the spectrum of Markarian~421.
However, the 4\,$\sigma$ excess at energies greater than 20~TeV in our data
favors a cutoff energy of $\sim$8~TeV,
at the upper end of the range previously reported from 
measurements at TeV energies.
\end{abstract}


\section{Introduction}

Markarian~421 (Mrk~421, J1104+3812) is a nearby BL Lacertae object ($z=0.031$)
and was the first extragalactic TeV gamma-ray source discovered
\citep{punch92}.
The TeV gamma-ray flux is variable, with flaring behavior
observed on time-scales of less than an hour \citep{gaidos96}.
Extensive measurements have been performed by 
several experimental groups based on the imaging Cherenkov 
technique~\citep{aharonian99_1,krennrich99,piron01}.
Multi-wavelength observations
support the Synchrotron--Self--Compton (SSC) mechanism
for the production of TeV gamma-rays from this source
~\citep[see, e.g.,][]{takahashi00,krawczynski01}.

TeV gamma-rays from extra-galactic
sources suffer absorption due to 
photon-photon interactions
with the inter-galactic infrared background
radiation~\citep{hikishov62,gould67,stecker92}.
According to recent measurements of 
the infrared background~\citep[see, e.g.,][and references therein]{hauser01} 
and predictions of the optical depth 
for TeV gamma-rays~\citep{primack99,dejager01,totani02},
gamma-rays at energies above 10~TeV from Mrk~421 
are expected to be suppressed,
since they interact with mid- to far-infrared photons of $\sim$100~$\mu$m.

Mrk~421 became active in 2000 and 2001,
especially at the beginning of 2001~\citep{boerst01}.
During this period,
northern hemisphere observers
measured the energy spectrum with good statistics
in the region from several hundred GeV to $\sim$10~TeV
and reported cutoffs at 3--6~TeV
\citep[]{krennrich01,aharonian02b}.
The cutoff energy is consistent with, or slightly smaller than, 
that measured for Mrk~501 during its flaring state 
in 1997~\citep{aharonian99_2,aharonian01}.
As Mrk~501 has a similar redshift ($z=0.034$) to Mrk~421,
this suggests the cutoffs may be due to infrared absorption of TeV gamma-rays.

We observed Mrk~421 during the 2001 high state
with the CANGAROO-II 10~meter telescope, 
at very large zenith angles of $\sim$70$^\circ$.
Similar observations have been reported by the Durham group
for Mrk~501 in the high state of 1997~\citep{chadwick99}.
For these observations, an effective collecting area
$\sim$10 times larger than that for observations near the zenith is
obtained, with an accompanying
increase in the gamma-ray energy threshold to $\sim$10~TeV.

\section{Observations and Analysis}

The observations were made with the CANGAROO-II 10~meter 
telescope~\citep{mori01,tanimori01},
located near Woomera, South Australia, Australia 
(136$^\circ$47$'$E, 31$^\circ$06$'$S).
The telescope consists of 114 segmented 
optical mirrors, each of 80~cm diameter~\citep{kawachi01}.
The camera contains 
552 half-inch photomultiplier tubes,
arranged at 0$^\circ$.115 intervals,
and covering a field of view of $\sim$3$^\circ$.

Mrk~421 was observed for ten nights in early 2001;
January 24, 26, 27, 30, 31 and February 1, and
March 1--4 (all dates in UT),
when the source was extremely active.
From the CANGAROO-II telescope site,
Mrk~421 culminates at a zenith angle of 69$^\circ$.3.
Approximately two hours observations were made per night.
OFF source data were taken with the right ascension suitably
offset.
An event trigger was registered when 3 individual pixels exceeded
a threshold of $\sim$2.5 photoelectrons. 
After rejecting data affected by clouds and 
those at zenith angles greater than 71$^\circ$.5,
14.34~hours ON source data and 16.65~hours OFF source data remain.
A software trigger was applied in order to reduce the effect of pixels
randomly triggered by the night sky background.
Pixels with pulse-heights of greater than $\sim$3.3 photoelectrons, 
pixel trigger times within 40~nanoseconds of the central value for the event, 
and three or more adjacent pixels were required.
Finally, four or more pixels surviving these cuts 
were required in each event.

Large zenith angle observations are well-suited to searching
for gamma-ray signals at higher energies,
as a much larger effective area 
can be achieved compared to observations near the zenith 
~\citep{sommers87, tanimori94},
though with a higher energy threshold.
From Monte Carlo simulations~\citep{okumura01,enomoto02a},
an effective area of $\sim5\times10^9~\mbox{cm}^2$ at $E=20$~TeV
was estimated 
for observations at 70$^\circ$, with the area
increasing to $\sim10^{10}~\mbox{cm}^2$ 
for higher energies.
A threshold energy 
(where the gamma-ray detection rate is maximized)
of $\sim$11~TeV 
was derived for a $E^{-3.0}$ spectrum.
This is an increase by a factor of $\sim$30 in comparison
with observations near zenith.
The energy threshold changes by $\sim\pm$1~TeV if the spectral
index is varied by $\pm$0.5.

The selection of the gamma-ray events is based on 
the parameterization of the elongated shape of 
the Cherenkov light image using the standard parameters:
$width$, $length$ (shape), $distance$ (location), $asymmetry$ (direction),
and $alpha$ (orientation angle)~\citep{hillas82,punch93,reynolds93}.
Instead of the conventional parameterization cuts,
we adopted the Likelihood method~\citep{enomoto02a, enomoto02b},
which has a higher efficiency of 
gamma-ray discrimination than the conventional parameterization 
technique.
The likelihood method uses a single parameter, 
$R_{prob} = Prob(\gamma)/[Prob(\gamma)+Prob(B.G.)]$,
where $Prob(\gamma)$ and $Prob(B.G.)$ are 
the probabilities for the event having been initiated by a
gamma-ray from the source or a background event, respectively. 
They are the products of individual probabilities for
$width$, $length$, and $asymmetry$, 
which are derived from the probability density functions,
including the energy dependence.
These functions were obtained
using gamma-ray simulations for the signal
and the observed OFF source events for the background.
$R_{prob}$ ranges from 0~to~1, 
and the probability of a gamma-ray origin for an event increases
as $R_{prob}$ becomes closer to 1,
though the gamma-ray acceptance does not significantly decrease
in the range of $R_{prob} \lesssim$~0.5.
We adopted a relatively loose cut of $R_{prob} >$~0.4 
with an additional requirement of $0^\circ.2<distance<1^\circ.1$.
With these cuts and a further cut excluding events with $alpha \geq 20^\circ$, 
86~\% of background events are rejected
while 63~\% of gamma-ray events are expected to be retained.

The resulting event distribution of 
$alpha$ is shown in Figure~\ref{fig:alpha}~(a) (left panel).
A clear excess over the background is apparent around the source direction. 
The excess is broadly distributed, up to $\sim$~30$^\circ$,
due to the deterioration of the pointing resolution,
caused by the shrinkage of the gamma-ray shower image.
This spread in $alpha$ distribution is
consistent with simulations,
as shown in the bottom panel of Figure~\ref{fig:alpha}.
The OFF source distribution was normalized to that of the ON source by the 
ratio of the number of the events in $alpha>40^\circ$ (0.88), 
which is consistent with the ratio of observation times (0.86), 
within statistical errors.
An excess of 298\,$\pm$\,52~events, 
with a significance of 5.7~$\sigma$
(calculated using the method of \citet{lima83})
was obtained in the region of $alpha < 20^\circ$.
For the confirmation of the detected signal, 
the conventional parameterization cuts of
0$^\circ$.2$<distance<$1$^\circ$.1, 0$^\circ$.06$<length<$0$^\circ$.18, and 0$^\circ$.03$<width<$0$^\circ$.14
were applied to the data,
and a signal of 286\,$\pm$\,55~events was obtained with 5.2~$\sigma$ 
significance.

\begin{figure}
\plotone{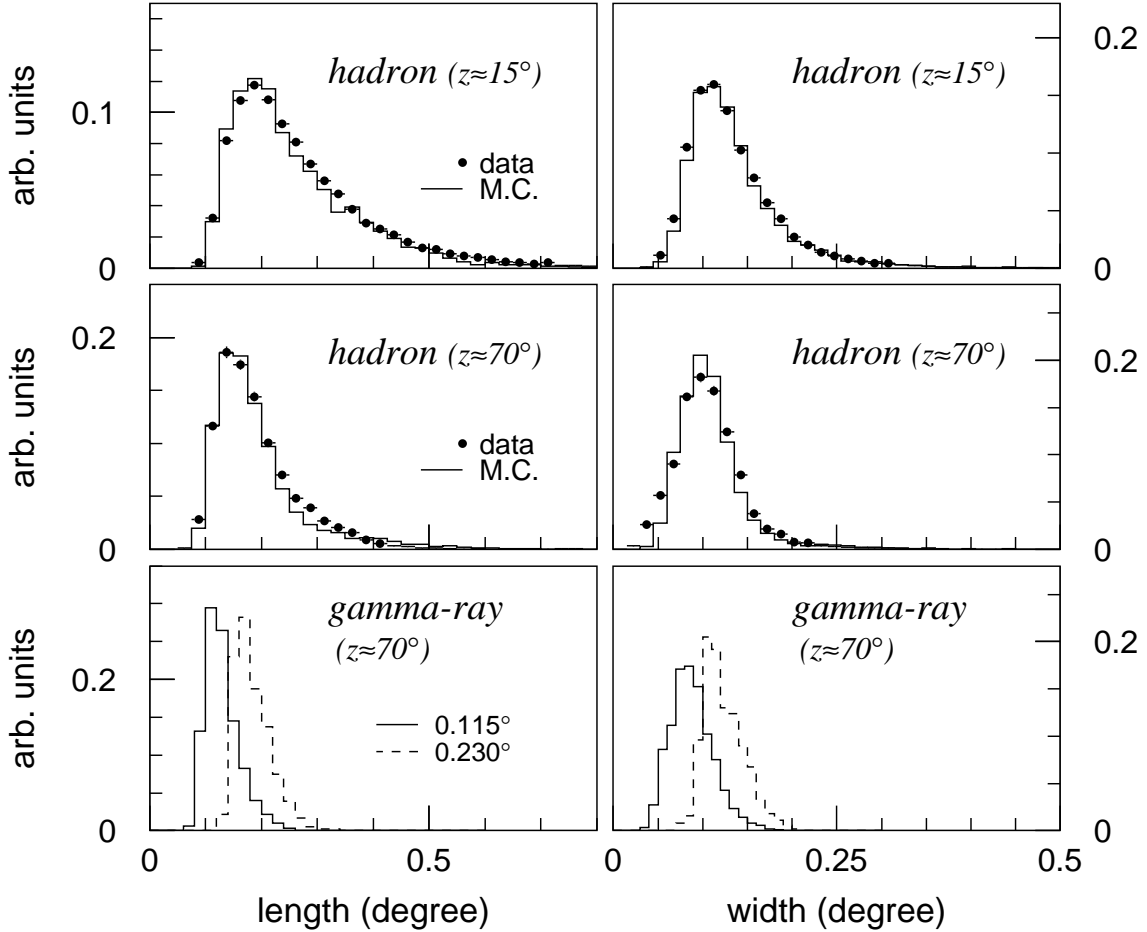}
\caption{Distributions of the image-centroid parameters
($length$ and $width$) observed at the large ($\sim$70$^\circ$) 
and small ($\sim$15$^\circ$) zenith angles.
In the upper and middle panels,
the observed background data (dots with error bars) 
and hadrons simulations (solid lines) are shown
for the large and small angles. 
In the bottom panels,
those of the gamma-ray simulations, with different 
camera pixel spacings (0$^\circ$.115 and 0$^\circ$.230), are shown 
for large angle observations.
\label{fig:para}}
\end{figure}

Since the observations were undertaken at
large zenith angles, $\sim$~70$^\circ$,
we carefully examined the data and the simulations 
in more detail:

1. The shrinkage of the shower image, 
which is problematic for large angle observations, 
was studied by a comparison between simulations and data
using the background events due to cosmic-ray hadrons.
Figure~\ref{fig:para} shows the 
imaging parameters $length$ and $width$, 
observed
at large ($\sim$70$^\circ$) and small ($\sim$15$^\circ$)
zenith angles, respectively.
The hadron simulations were made 
using the CORSIKA code (version 6.004)~\citep{corsika},
considering the cosmic-ray abundance in the TeV region~\citep{mohanty98}.
The resultant distributions of the simulations agree with the
data for both small and large angles.

It is also noted that our high resolution imaging camera, 
which has a pixel spacing size of 0.115$^\circ$, 
helped to separate the smaller images of gamma-ray events 
from those of background events.
The expected $length$ and $width$ distributions of the gamma-ray events,
simulated with the spacing size of 0$^\circ$.115 and 0$^\circ$.230,
are shown in the bottom panels of Figure~\ref{fig:para}.
With the larger pixel size,
the reconstructed image size increases and becomes more similar 
to those of hadrons, with an 
estimated $\sim$50~\% decrease in the separation efficiency.
Although gamma-ray detection is still possible with the larger spacing,
the higher resolution imaging camera 
is more advantageous for large zenith angle observations.

\begin{figure}
\plotone{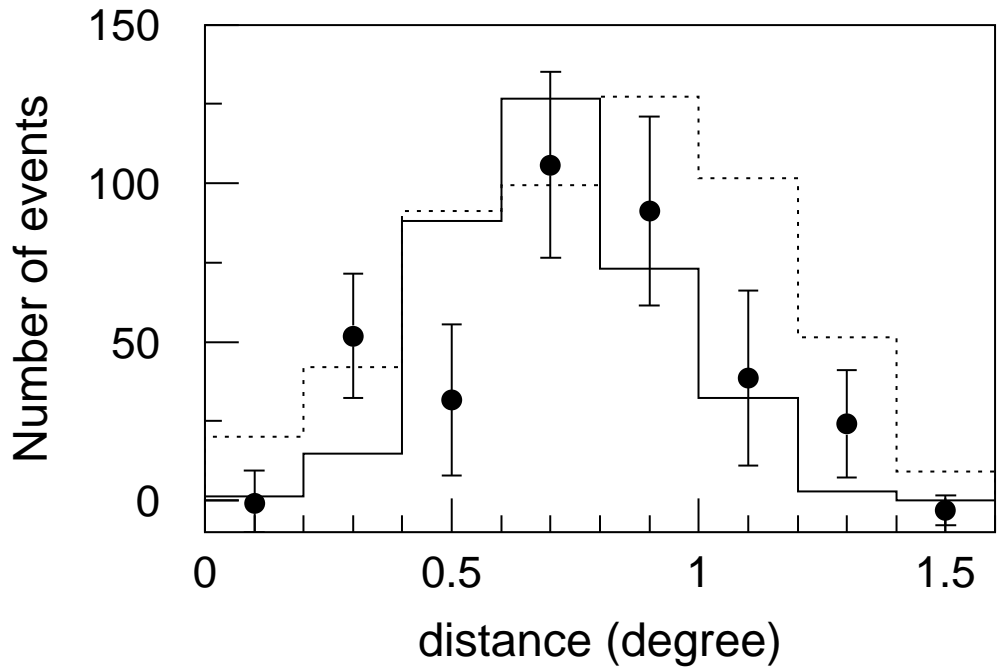}
\caption{Distributions of the image shape parameter $distance$, 
after subtracting normalized OFF-source data from ON-source data
(circles with error bars), 
gamma-ray simulation (solid line) 
and hadron background (dotted line).
\label{fig:distance}}
\end{figure}

2. The $distance$ distribution of the gamma-ray selected events
was compared to those from simulations.
The location of Cherenkov images due to gamma-ray  
cascades in the field of view has a particular distribution,
while those due to hadron showers are uniformly distributed.
For large zenith angle observations in particular,
as the observed distances of gamma-ray shower images decrease,
there is a substantial difference with the background distribution.
Figure~\ref{fig:distance} shows 
the $distance$ distribution of the gamma-ray selected events,
which is obtained by subtracting the OFF source distribution from 
the ON source distribution 
after the likelihood and $alpha$ cuts were applied.
The resulting distribution has a clear peak around 0$^\circ$.7
from the source direction, 
which agrees reasonably well with simulations
and differs from that of the cosmic ray background,
which provides additional confirmation
of the detection of TeV gamma-rays.

3.  The ``standard candle'' at TeV energies, the Crab nebula, 
was observed at relatively large zenith angles of $\sim$55$^\circ$
in November and December 2000,
and the gamma-ray flux was measured for the confirmation of 
the analysis method and the estimation of the systematic error in
the energy scale. 
Using the same analysis technique as that used for Mrk~421,
the differential energy spectrum was derived 
over the energy range from 2~TeV to $\sim$20~TeV~\citep{itoh02},
which agrees well with 
other experiments~\citep{tanimori98,aharonian00,krennrich01},
within a $\sim$15~\% error in the energy scale.

These consistencies provide 
robust supporting evidence for the detection of 
$E>10$~TeV gamma-rays from Mrk~421.

\section{Discussion}

\begin{figure}
\plotone{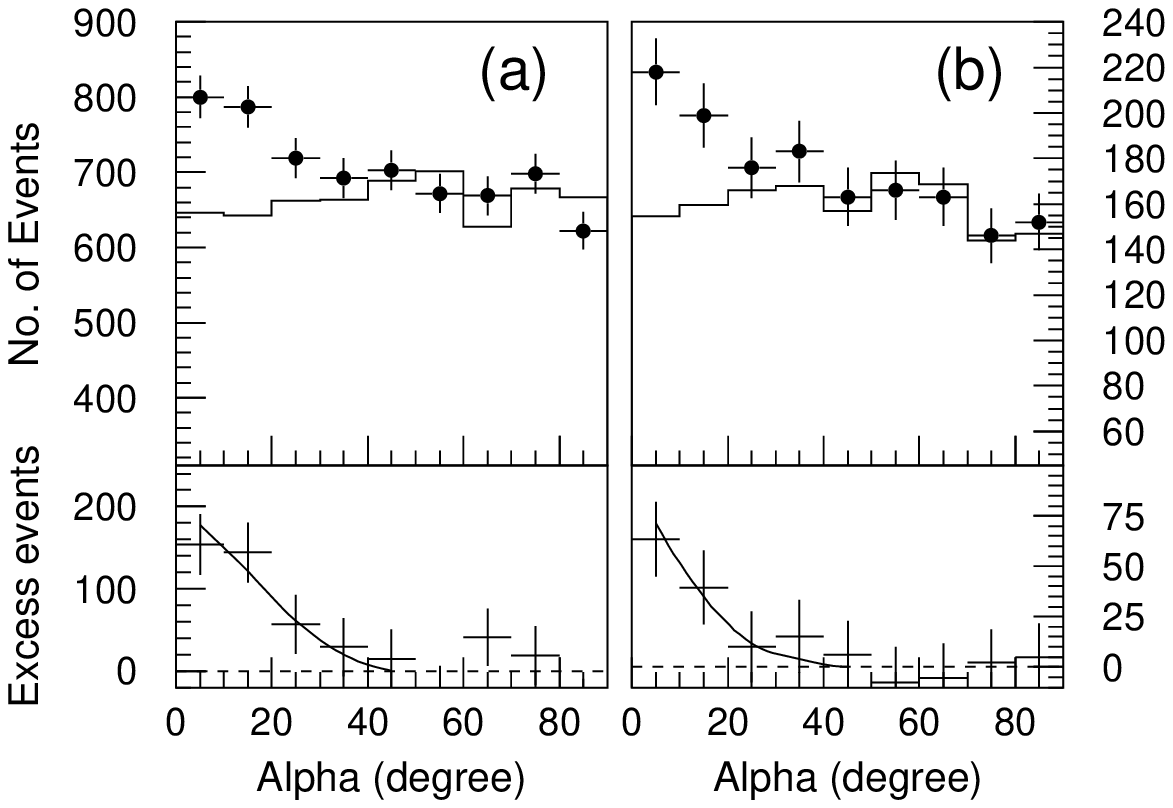}
\caption{Image orientation angle ($alpha$) distributions for
gamma-ray--like events with respect to the direction to Markarian 421.
The left figure (a) shows the distributions for all energies, and 
the right figure (b) for those with reconstructed energies above
 20~TeV.
In the upper panel,
filled circles with error bars (statistical only) and solid lines 
are for the ON and OFF source data, respectively.
The lower panel shows the excess events of the ON source 
above the background (OFF source) level.
The solid curves show the expected spread of 
gamma-ray events in the $alpha$ distribution from simulations.
\label{fig:alpha}}
\end{figure}

\begin{figure}
\plotone{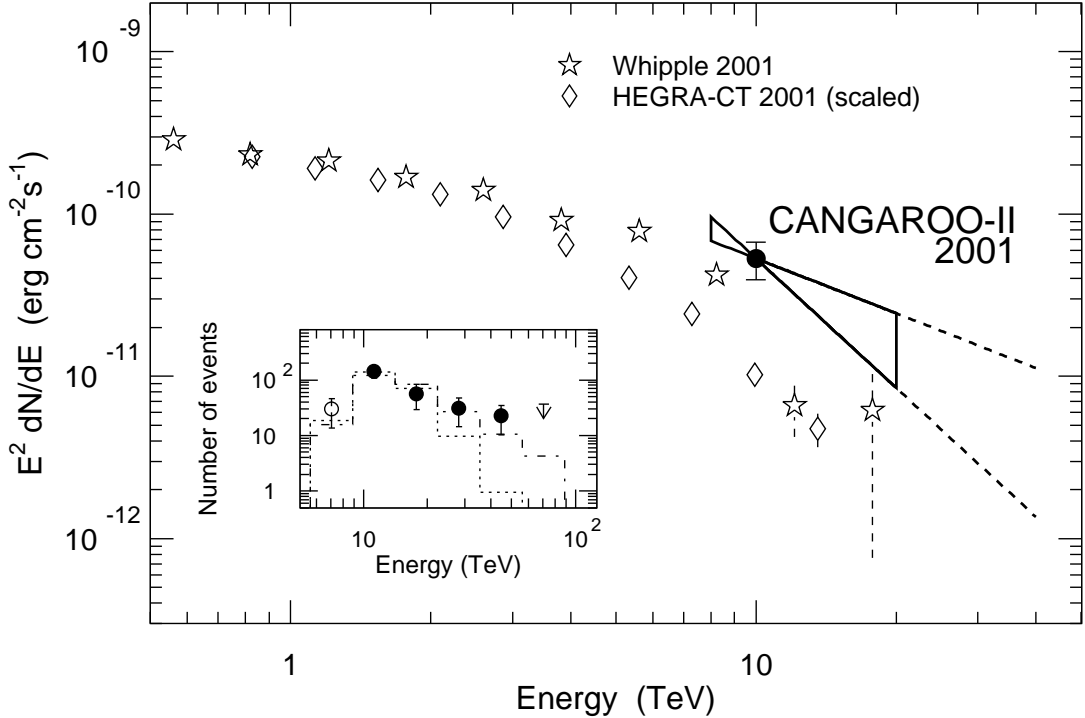}
\caption{
The observed gamma-ray fluxes (main panel) and 
the energy spectrum of gamma-ray events (inserted panel).
In the inserted panel, data are represented by circles with error bars,
with a 2~$\sigma$ upper limit plotted at the highest energy.
Best-fit spectra for a
power-law ($E^{-4.0}$ ; dot-dashed line) and a
cut-off ($E^{-1.9}\exp(-E/\mbox{4~TeV})$ ; dotted line) 
are shown (see text for details).
The data shown with the filled circles were used for 
the spectral shape fitting.
In the main panel,
the measured flux under the assumption of a power-law spectrum is
shown with error bars and the area corresponding 
to statistical errors of $\pm$1~$\sigma$.
Whipple~\citep{krennrich01} and HEGRA-CT~\citep{aharonian02b} spectra
measured in similar periods are also shown.
The fluxes plotted for the HEGRA-CT group have been
scaled in order to normalize it to the Whipple flux at 1~TeV.
\label{fig:energyflux}}
\end{figure}

Figure~\ref{fig:energyflux} (inserted panel) shows 
the raw energy spectrum of the observed gamma-ray events from Mrk~421.
The gamma-ray energy was assigned from 
the pulse-height sum of the individual pixels,
using a relation obtained from the simulations.
This method is similar to that described in \citet{mohanty98},
and an energy resolution of $\sim$31~\% is estimated.
The excess events are distributed in the energy range 7--45~TeV,
however one must take care of the spill-over effect from the lower
energies due to the finite energy resolution.
In order to take this effect into account,
simulated gamma-ray spectra, with
the spectral indexes and cutoff energies varied,
were compared to the data and the observed spectral parameters were 
determined from the values which minimized the value of $\chi^2$.
With the assumption of a power-law spectrum, 
the differential flux was fitted by

$$
\frac{dN}{dE} = (3.3\,\pm\,0.9_{stat.}\,\pm\,0.3_{syst.}) \times 10^{-13}\ 
\left(\frac{E}{10~\mbox{TeV}}\right)^{-(4.0\,^{+0.9}_{-0.6}\,_{stat.}\,\pm\,0.3_{syst.})}
\ ~~~~\mbox{ph.}/\mbox{cm}^{2}/\mbox{sec}/\mbox{TeV}
$$

\noindent
with  $\chi^2$=2.5/2~$d.o.f.$
The cut dependence on $R_{prob}$ and $alpha$ parameters, and 
the trigger conditions in the simulation, were considered
as sources of the systematic uncertainties.
The systematic errors giving rise to uncertainty in the energy scale
such as Cherenkov photon scattering in the atmosphere
are not included here, but are considered in more detail later.
The derived spectrum is steeper than those observed at lower TeV energies.
The spectral shape was tested with a cutoff spectrum of
$E^{-1.9}\exp(-E/4\mbox{TeV})$,
as was derived from the measurements 
by the Whipple and HEGRA-CT groups, 
with the spectral index being the hardest one
observed during the strong flaring period~\citep{aharonian02b,krennrich02}.
The fitting result did not improve compared to
that with the power-law assumption ($\chi^2$=5.0/3~$d.o.f.$),
as an excess of events above 20~TeV is apparent,
as shown in Fig~\ref{fig:alpha}~(b).
An excess of 103\,$\pm$\,26 (4.0~$\sigma$) was observed with $alpha<20^\circ$,
while 11~events are expected for the cutoff spectrum, based
on an estimation using the event ratio between 10--20~TeV and over 20~TeV.
However, if a cutoff energy of 8~TeV is assumed,
the consistency with the data becomes better 
(48 events expected for $E^{-1.9}\exp(-E/8\mbox{TeV})$).
This cutoff energy is at the high end of the range allowed for 
Mrk~501~(\citealt{aharonian99_2}, see also \citealt{aharonian01}).
Since these two AGNs have similar redshifts,
the cutoff energies in both spectra are expected to be similar,
assuming the attenuation is predominantly due to infrared
absorption.
As there is only a 2\,$\sigma$ difference between our observations
and this prediction, 
our result falls in the acceptable range of the absorption 
hypothesis due to the cosmic infrared background.

Figure~\ref{fig:energyflux} (main panel) shows the measured energy flux,
assuming the power-law spectrum.
Data for the Whipple~\citep{krennrich01} 
and HEGRA-CT groups~\citep{aharonian02b},
observed during a similar period of the flaring state
(January--March 2001) are also shown.
The observation periods were not exactly the same
and the source varied significantly during this high state,
therefore the absolute fluxes are expected to differ at some level.
The absolute flux level determined from the CANGAROO-II data is within
the observed range of the flux variation 
reported by the Whipple group~\citep{krennrich02},
and the spectral slope around 10~TeV
is consistent with that of these two groups, 
supporting the roll-over from
the flatter spectrum measured at lower energies.

For large zenith angle observations, 
a large uncertainty in the energy scale,
due to the absorption of Cherenkov photons in the atmosphere,
is inevitable.
Only Rayleigh scattering was considered in the simulation code 
to avoid over-estimating the gamma-ray energies.
The inclusion of Mie scattering and ozone absorption 
would affect the energy scale by $\sim$30~\% and $\sim$3~\%,
respectively, based on numerical estimations
using the program code of \citet{kneizys96}.
We stress that these effects increase the energy scale.
The use of the  ``flat-Earth'' approximation 
for the atmosphere in the simulations requires a $\sim$6~\% correction which
has already been taken into account in the discussion above.

The measurement of spectra at large zenith angles
was verified by observations of the Crab nebula
up to the zenith angles of $\sim$55$^\circ$,
although calibration using the Crab nebula at the same zenith angles
as the Mrk~421 observations ($\sim$70$^\circ$) is 
unfortunately impractical with the current instrumental sensitivity.
The strong gamma-ray emission of Mrk~421 
($\sim$~3 times that of Crab nebula) 
enabled us to detect the source in only 14~hours.
In order to 
detect the Crab nebula at the same significance level,
more than 150~hours observations would be required.

In summary,
owing to the large effective area 
and the high resolution performance of the Cherenkov imaging camera,
E$>$10~TeV gamma-rays from Mrk~421 were detected 
at a high confidence level 
at zenith angles of $\sim$70$^\circ$
with 14~hours of observations.
The derived spectrum in the region of 10--30~TeV 
is steeper than that around 1~TeV, 
which supports the cutoff spectrum of 
Mrk~421 measured in the 0.2--10~TeV range by other groups.
The excess observed above 20~TeV
is strongly suggestive of a higher cutoff energy, $\sim$8~TeV,
compared to the lower energy observations.
These observations confirm, with the support of detailed simulations,
the viability of the large zenith angle technique.
Large zenith angle observations provide a unique method 
of measuring the spectrum in the important energy range above 10~TeV
with a relatively short observation time.

\acknowledgments

The authors thank F.~Krennrich and D.~Horns for kindly 
providing flux data.
This project is supported by a Grant-in-Aid for
Scientific Research of Ministry of Education,
Culture, Science, Sports and technology of Japan
and Australian Research Council.
The receipt of JSPS Research Fellowships is also acknowledged.
We thank the DSC Woomera
for their assistance in constructing the telescope.

\epsscale{1.0}

\end{document}